# PLANT MODEL GENERATION FROM EVENT LOG USING PROM FOR FORMAL VERIFICATION OF CPS *

**Midhun Xavier**
Lulea Tekniska Universitet
Lulea, Sweden
`midhun.xavier@ltu.se`

**Victor Dubinin**
Penza State University
Penza, Russia
`dubinin.victor@gmail.com`

**Sandeep Patil**
Lulea Tekniska Universitet
Lulea, Sweden
`sandeep.patil@ltu.se`

**Valeriy Vyatkin**
Lulea Tekniska Universitet, Aalto University
Espoo, Finland
`vyatkin@ieee.org`

## ABSTRACT

This paper introduces the concept of plant model generation from the recorded traces of events using the process mining technique. The event logs are obtained by visually simulating a simple distributed manufacturing system using the OPC UA communication protocol. The process discovery alpha algorithm is used to extract the process model in Petri net format. The system behavior represented in terms of Petri net is then fed to construct the ECC of the basic function block in compliance with IEC 61499 standard using the proposed notation. Finally, the formal verification of the closed-loop system is done with the help of a tool chain that consists of fb2smv converter, symbolic model checker NuSMV, and other tools for representing counter-examples.

## 1 Introduction

The cyber-physical system (CPS) is quite popular in today's ever-evolving industry. Identifying the scenarios where and how it might fail in a distributed automation system consisting of intelligent mechatronic components is arduous. Running the system in real is considered to be expensive because some processes may cause severe damage to the system. Therefore, we can construct a simulation model of the real system using IEC 61499 standard [1] and see how it behaves, but there are intermittent errors which do not show up in simulation but can appear in real behavior. This simulation does not prove the absence of bugs, hence formal verification [2] could be used as a better approach to verify the correctness and safety properties of the industrial automation systems. Compared to simulation, formal verification [3] helps to computationally explore the behavior of the system thoroughly. Formerly, the developers used to construct plant models manually but this paper introduces a concept of plant model generation from event logs using process mining technique. The process mining is used for extracting the process models from the recorded behavioral traces of the system. The Petri net model of the system's behavior is derived with the help of ProM tool using an alpha algorithm [4]. The generated modular formal model of the closed-loop system is transformed to the model of uncontrolled plant behavior extended with non-determinism and its verification can be done with the help of the comprehensive tool-chain [5].

## 2 Case study

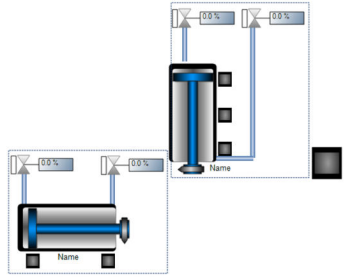

Figure 1: HMI representation of Two Cylinder system

In this experiment we use the Two-cylinder system example. The major components in the simulation model are the vertical and horizontal cylinders. The first component is a vertical cylinder which moves in an upward or downward direction. Whenever a work-piece is detected by the sensor under the drill, it moves downward and starts drilling. Once it completes drilling, it moves upwards and rests at the home position. Second one is the horizontal cylinder. When a work-piece is loaded onto the system, the horizontal cylinder pushes the item to the other end for drilling. Simulation model of the system is developed using Nxt control's Human Machine Interface (HMI) . The HMI representation of the system is shown in the figure 1. The event log is constructed by running the visual simulation model on nxtStudio using OPC UA communication protocol.We used the UaExpert tool as an OPC UA client to record the events and data from function blocks. The recorded event log in CSV format is then used for plant model implementation.

## 3 Plant modelling and Closed-loop verification of the system

### 3.1 Proposed Solution

The plant modelling and closed-loop verification of the system can be done using the workflow diagram shown in figure 2. There are several methods to construct the event log. The various process scenarios of the system are constructed by manually triggering the horizontal cylinder's 'EXT' actuator signal. The generated trace file in the CSV format is fed to the ProM tool. The ProM tool helps to pre-process the data and it constructs Petri net using an Alpha algorithm. The Petri net is then given to implement Finite State Machine(FSM) by using reachability analysis. This FSM is used for the implementation of a function block interface and an execution control chart (ECC) of the plant model with the help of proposed transformation notation. The developed function block model of the plant is then connected with the controller function block model in closed-loop and it is given for formal verification using a tool-chain.

### 3.2 Petri net generation using ProM tool

The event log recorded by running visual simulation of the two-cylinder system is shown in the figure 3 a. This event log contains different kinds of processes related to horizontal cylinder movement, i.e. it consists of different motion scenarios between the START and END position. The structure of the event log contains processId, timestamp, component and action. The processId is the unique id for each process scenario. Timestamp is the time at which the event has occurred. The component describes the part of the system, i.e. whether it is a horizontal or vertical cylinder and finally the action explains which event or condition has triggered. The event log related to the horizontal cylinder is extracted and is then fed to the ProM tool, which converts the event log in CSV format to eXtensible Event Stream (XES) format. The ProM tool has several plugins for creating Petri nets. One of the popular miners is the alpha algorithm, which takes the event log as input for producing Petri net. The developed Petri net for horizontal cylinder is shown in figure 3 b. This Petri net has the overall behavior of the horizontal system. In order to implement the horizontal cylinder's plant model, we need to apply transformation which is explained in the following section.

### 3.3 Plant model construction from Petri net

The event log used to construct the Petri net is recorded based on the horizontal cylinder behavior so the model created from this Petri net is called Reference Model of Control/Sensor Events Sequencing (RMCSES) and this will be a complete (full) model of the low-level functioning of the system because all signals are taken into account. However, the language generated by this formal model is wider than the set of event sequences presented in the event log. In addition, time delays in the Plant model (i.e., the time of occurrence of sensory signals after the issuance of their preceding control signals) are represented by spontaneous transitions, which introduces a certain non-determinism to the Plant model. Thus, the Plant model will include not only the nominal behavior under a certain control system, but also a wider (including undesirable) behavior. In addition, using the approach proposed in this paper, the problem of verification of the conformity of a (new) control system to be certified with the RMCSES can be solved. The figure 4 a, shows the transformation of the RCM interface to the interface of the basic function block of the plant model. The function block interface obtained after the transformation is shown in figure 4 b.

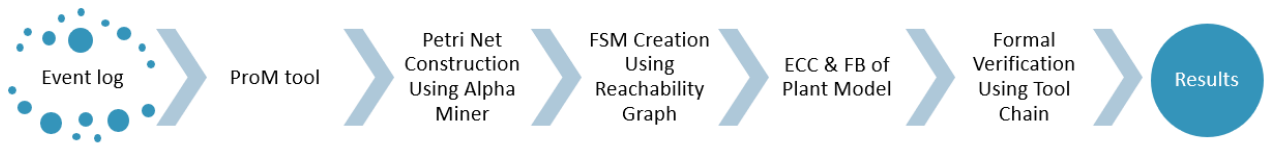

Figure 2: Workflow diagram

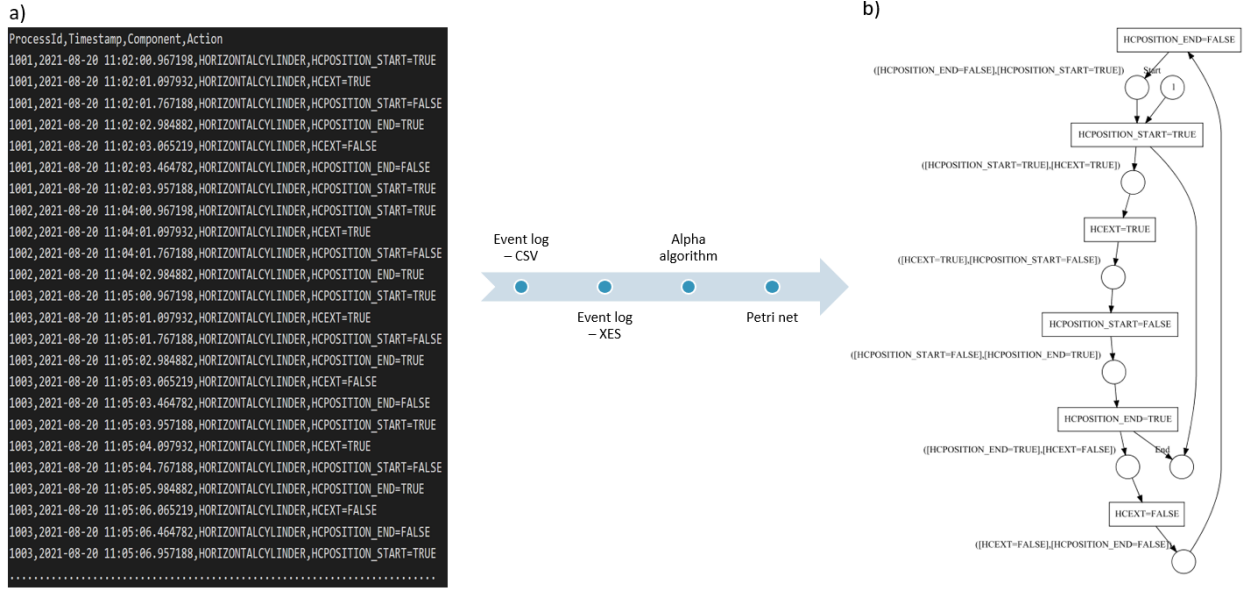

Figure 3: a) Event log; b) Petri net.

The reachability graph generated using PIPE tool is shown in figure 5 a, which takes Petri net as input. Also, we need to provide the initial markings and it is better to remove the Start and End states prior to reachability analysis. The ECC of the plant model is implemented by applying transformation (figure 5 b) on FSM extracted from Petri net. The sensor signal present on any transition of FSM is transformed into a non-deterministic transition (NDT) and the sensor signal is produced as an output event after this non-deterministic transition. The developed ECC of the plant model is shown in figure 5 c.

### 3.4 Formal verification of closed-loop system

The derived plant model FB from the event log is connected in a closed-loop with a controller for formal verification using tool-chain [5]. The closed-loop composite IEC 61499 FB is converted to SMV code using fb2smv converter and formal verification is done with the help of symbolic model checker NuSMV. The following CTL specification is verified :

```
-- specification  G !( FV_HC_CFB_inst.HC_CONTROLLER.HOME = TRUE & FV_HC_CFB_inst .
    HC_CONTROLLER.END = TRUE)  is  true
```

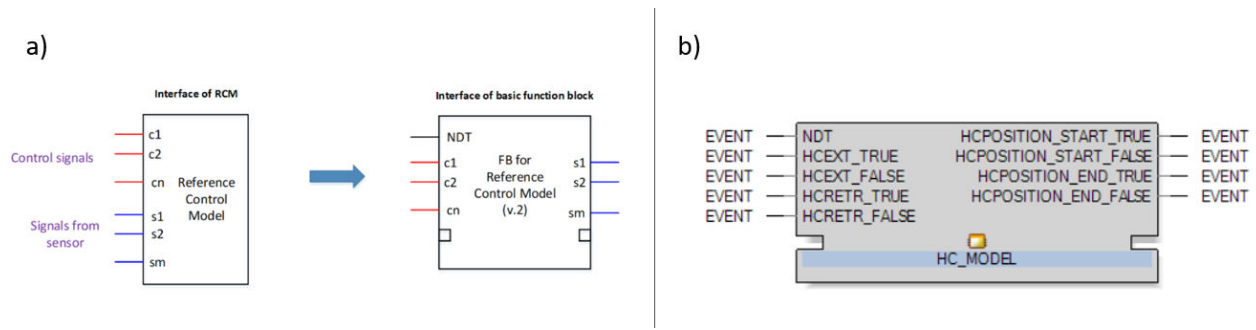

Figure 4: a) Transforming the Interface; b) FB Interface Horizontal Cylinder.

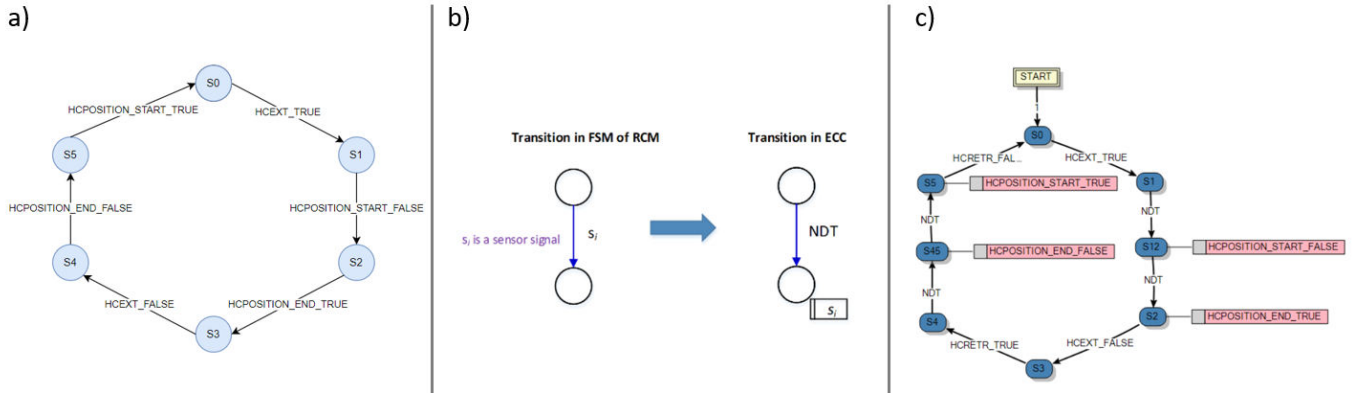

Figure 5: a) FSM extraced from Petri net; b) Transforming the State Chart; c) ECC of Horizontal Cylinder.

The following specification verifies that HOME and END plant sensors never be TRUE simultaneously.

## 4 Conclusion and Future plan

We implemented the plant model of the horizontal cylinder with the help of event log using process mining technique. The developed model of the plant is promising and its formal verification done with the help of various CTL or LTL specifications. Here, we recorded the event log by connecting the controller, but developing all possible trace scenarios could be the next step in future. The proposed method is fast for small examples but scalability and fastness of this approach in real examples needs to analyze and see how it behaves. Automatic generation of monitor and controller from event logs can added as a future work.

## Acknowledgments

This work was sponsored, in part, by the H2020 project 1-SWARM co-funded by the European Commission (grant agreement: 871743).